\documentclass[journal,10pt]{IEEEtran}

\usepackage{amssymb}
\usepackage{amsmath}
\usepackage{amsthm}
\usepackage{amsfonts}
\usepackage{accents}
\usepackage{multirow}
\usepackage{array}
\usepackage{multirow}
\usepackage{graphicx}
\usepackage{epstopdf}
\usepackage{float}
\usepackage{pgfplots}
\pgfplotsset{compat=1.8}
\usepackage{epstopdf,epsfig}

\ifCLASSOPTIONcompsoc
  \usepackage[caption=false,font=normalsize,labelfont=sf,textfont=sf]{subfig}
\else
  \usepackage[caption=false,font=footnotesize]{subfig}
\fi

\usepackage{booktabs}

\definecolor{mycolor1}{rgb}{0.00000,0.20000,0.60000}%
\definecolor{mycolor2}{rgb}{0.00000,0.40000,0.80000}%
\definecolor{mycolor3}{rgb}{0.40000,0.60000,1.00000}%
\definecolor{mycolor4}{rgb}{0.00000,0.60000,1.00000}%
\definecolor{mycolor5}{rgb}{1.00000,0.40000,0.60000}%
\definecolor{mycolor6}{rgb}{1.00000,0.20000,0.40000}%
\definecolor{mycolor7}{rgb}{1.00000,0.00000,0.20000}%
\definecolor{mycolor8}{rgb}{0.60000,0.20000,0.00000}%

\begin{document}
\title{Tomlinson-Harashima Cluster-Based Precoders for Cell-Free MU-MIMO Networks}% \vspace{-0.05em}}

\author{André R. Flores$^1$, Rodrigo C. de Lamare$^{1,2}$
and Kumar Vijay Mishra$^3$ \\
$^1$Centre for Telecommunications Studies, Pontifical Catholic
University of Rio de Janeiro, Brazil \\
$^2$Department of Electronic Engineering, University of York,
United Kingdom \\
$^3$United States DEVCOM Army Research Laboratory, Adelphi, MD 20783 USA \\
Emails: \{andre.flores, delamare\}@cetuc.puc-rio.br, kvm@ieee.org
\vspace{-0.01cm}
}

% *** make the title area ***
\maketitle

\begin{abstract}
Cell-free (CF) multiple-input multiple-output (MIMO) systems generally employ linear precoding techniques to mitigate the effects of multiuser interference. However, the power loss, efficiency, and precoding accuracy of linear precoders are usually improved by replacing them with nonlinear precoders that employ perturbation and modulo operation. In this work, we propose nonlinear user-centric precoders for CF MIMO, wherein different clusters of access points (APs) serve different users in CF multiple-antenna networks. Each cluster of APs is selected based on large-scale fading coefficients. The clustering procedure results in a sparse nonlinear precoder. We further devise a reduced-dimension nonlinear precoder, where clusters of users are created to reduce the complexity of the nonlinear precoder, the amount of required signaling, and the number of users. Numerical experiments show that the proposed nonlinear techniques for CF systems lead to an enhanced performance when compared to their linear counterparts. 
\end{abstract}


% \fi
%\IEEEpeerreviewmaketitle

\section{Introduction}
Coordinated base stations (BSs) have been deployed worldwide to establish cellular network services. However, wireless applications are evolving constantly with an increasing demand for more resources \cite{Tataria2021,Giordani2020,mmimo,wence}. For high throughput and quality-of-service required for future networks, it is desired to further densify BSs. However, this approach is impractical. As an alternative, cell-free (CF) multiple-input multiple output (MIMO) systems have emerged as a potential solution to improve the performance and satisfy throughout requirements of future wireless networks \cite{Ammar2022,Elhoushy2021}. 

Compared to conventional BS-based networks, CF MU-MIMO systems employ multiple APs distributed geographically over the area of interest. A central processing unit (CPU), which may be located at the cloud server, coordinates the APs. The distributed deployment of CF networks yields higher coverage than the BSs with collocated antennas \cite{Attarifar2020}.  In addition, CF multiuser MIMO (MU-MIMO) has been shown to provide increased throughput per user \cite{Yang2018,Elhoushy2021b} as well as better performance in terms of energy efficiency \cite{Ngo2018,Zhang2019,Jin2021}. 

Further, CF MU-MIMO employs the same time-frequency resources to provide service to multiple users as BS-based systems. To avoid the multiuser interference (MUI) in the downlink, a precoder is often implemented at the transmitter. Prior works on CF MU-MIMO have focused on linear precoding techniques such as matched filter (MF), zero-forcing (ZF) \cite{Nayebi2017}, and minimum mean-square error (MMSE) \cite{Bjoernson2020} techniques. However, it is well-known that nonlinear precoders \cite{Zu2014,rmbthp,rsthp} have the potential to outperform their linear counterparts \cite{Joham2005,lrcc,Palhares2020,rsrbd,cesg,cfrs}.

State-of-the-art in CF MU-MIMO systems has proposed network-wide (NW) precoders \cite{Ngo2017,Nayebi2017,Bjoernson2020,Nguyen2017} but these techniques entail a very high signaling load. Moreover, NW approaches demand high computational complexity because they require the inversion of a matrix whose size increases with the number of APs and users. To mitigate this problem, NW precoders that employ APs and user clusterization have been proposed \cite{Buzzi2020} for lower computational complexity and signaling load. For instance, in \cite{Palhares2020} the number of APs is curtailed to reduce the signaling load. In \cite{Bjoernson2020a}, scalable MMSE combiners and precoders are developed. Very recently, a regularized ZF precoder based on subsets of user was proposed in \cite{Lozano2021} to judiciously use the available resources. 

Unlike previous works \cite{albreem2021overview,Palhares2020,cfrs}, we propose nonlinear precoding techniques for CF MU-MIMO systems. The proposed techniques are based on the well-established Tomlinson-Harashima precoder (THP) \cite{Fischer2002}, which may be interpreted as the transmit analog of the successive interference cancellation (SIC) employed at the receiver \cite{spa}. Essentially, THP employs a nonlinear modulo operation that reduces the power penalty associated with the linear precoders thereby enhancing the overall performance. Additionally, a cluster-based approach is devised based on a user selection matrix, resulting in a user-centric nonlinear precoder and addressing the gap in nonlinear structures for cluster-based precoders in CF networks. %The main contributions of this paper can be summarized as follows. A TH precoding technique for CF systems is proposed. AP selection is performed and cluster-based precoders are developed based on a user selection matrix. 
The resulting precoder is sparse and its complexity is reduced by employing clusters of users, thereby reducing the amount of signaling and the computational load. Our numerical experiments show that the TH precoding techniques outperform their linear counterparts.  

The rest of this paper is organized as follows. In the next section, we describe the ssytem model of the CF MU-MIMO communications. We derive the proposed cluster-based nonlinear precoding techniques in Section III. We introduce the metric to evaluate the performance of the proposed precoders in Section IV. We validate our model and methods via numerical experiments in Section V. We conclude in Section VI.  

\section{System Model}
Consider the downlink of a CF MIMO system, where $N$ geographically distributed APs serve $K$ users equipped with a single omnidirecitonal antenna. A central processing unit (CPU) located at the cloud server is connected to the APs. The data are transmitted over a flat-fading channel $\mathbf{G}\in\mathbf{C}^{N\times K}$. The $(n,k)$-th element of matrix $\mathbf{G}$ is the channel coefficient between the $n$-th AP and $k$-th user, i.e.,  $g_{n,k}=\sqrt{\zeta_{n,k}}h_{n,k}$, where $\zeta_{n,k}$ is the large-scale fading coefficient that models the path loss and shadowing effects, and $h_{n,k}$ represents the small-scale fading coefficient. The coefficients $h_{n,k}$ are modeled as independently and identically distributed (i.i.d.) random variables with complex Gaussian distribution $\mathcal{CN}\left(0,1\right)$.

Denote the transmit signal by $\mathbf{x}\in \mathbb{C}^{N}$, which obeys the transmit power constraint  $\mathbb{E}\left[\lVert\mathbf{x}\rVert^2\right]\leq P_t$, where $\mathbb{E}[\cdot]$ denotes the statistical expectation. Then, the $K \times 1$ received signal vector is
\begin{equation}
\mathbf{y}=\mathbf{G}^{\text{T}}\mathbf{x}+\mathbf{n},
\end{equation}
where $(\cdot)^T$ is the conjugate transpose and $\mathbf{n}\in \mathbb{C}^{K}$ is the additive white Gaussian noise (AWGN) that follows the distribution $\mathcal{CN}\left(\mathbf{0},\sigma_n^2\mathbf{I}\right)$.

The system employs the time division duplexing (TDD) protocol and therefore the channels can be estimated employing the channel reciprocity property and pilot training \cite{Vu2007}. After receiving the pilots, the CPU computes the channel estimate $\mathbf{\hat{G}}^{\text{T}}=\left[\mathbf{\hat{g}}_1,\mathbf{\hat{g}}_2,\cdots,\mathbf{\hat{g}}_k\right]^{\text{T}}\in\mathbb{C}^{K\times N}$, whose $(n,k)$-th element is
\begin{equation}
    \hat{g}_{n,k}=\sqrt{\zeta_{n,k}}\left(\sqrt{1-\sigma_e^2}h_{n,k}+\sigma_e\tilde{h}_{n,k}\right),
\end{equation}
where $\hat{g}_{n,k}$ is the channel estimate between the $n$-th AP and the $k$-th user; $\tilde{h}_{n,k}$ are i.i.d complex Gaussian random variables that follow the distribution $\mathcal{CN}\left(0,1\right)$ (independent from $h_{n,k}$) and model the errors in the channel estimates; and $\sigma_e$ represents the quality of the channel state information (CSI). The error affecting the channel estimate $\hat{g}_{n,k}$ is $\tilde{g}_{n,k}=\sigma_e\sqrt{\zeta_{n,k}}\tilde{h}_{n,k}$.

\section{Proposed Cluster-Based Nonlinear Precoders}

To enhance the performance of the system while reducing the signaling load and computational complexity of NW precoders, we propose cluster-based nonlinear precoders. To this end, we form clusters of APs and users. These clusters are defined based on the large-scale channel coefficients given by $\zeta_{n,k}$. Since only small subsets of APs transmit the most relevant signals for reception, the contribution of the remaining APs is not significant and the transmission over such APs is avoidable. The upshot of this technique is that we discard the APs whose processing is cost-ineffective to reduce the signaling load.
\subsection{AP selection}
The signaling load is brought down by taking into account that each user is served only by a reduced cluster of APs. Consider the pre-fixed scalar $L$ that denotes the number of APs that are going to be selected. Then, for the $k$-th user, the $L$ APs with the largest large-scale fading coefficient are selected and gathered in the set $\mathcal{A}_k$. In this sense, we employ the equivalent channel estimate $\bar{\mathbf{G}}^{\text{T}}=\left[\mathbf{\bar{g}}_1,\mathbf{\bar{g}}_2,\cdots,\mathbf{\bar{g}}_k\right]^{\text{T}}\in \mathbb{C}^{K \times N}$, which is a sparse matrix with the $(n,k)$-th element as
\begin{equation}
    \bar{g}_{n,k}=\begin{cases}
\hat{g}_{n,k},&n\in \mathcal{A}_k,\\
0, &\text{otherwise.}
\end{cases}
\label{sparse effective channel}
\end{equation}

\subsection{Sparse TH precoder}
Using \eqref{sparse effective channel}, we compute a sparse TH precoder (TH-SP), which defines how the symbols are transmitted by the selected APs. The conventional THP employs three different filters \cite{chen2019low}: feedback filter $\mathbf{B}\in\mathbb{C}^{K \times K}$, feedforward filter $\mathbf{F}\in\mathbb{C}^{N\times K}$, and a scaling matrix $\mathbf{C}\in \mathbb{C}^{K \times K}$ \cite{Zu2014}. The feedback filter $\mathbf{B}$ deals with the multiuser interference (MUI)
by successively subtracting the interference of previous symbols from the current symbol and, therefore, is a matrix with a lower triangular structure. The feedforward filter $\mathbf{F}$ enforces the spatial causality. The scaling matrix $\mathbf{C}$ assigns a weight to each stream and is, therefore, a diagonal matrix. Depending on the position of matrix $\mathbf{C}$, two different THP structures have been suggested: the centralized THP (cTHP) implements the scaling matrix at the transmitter side (at the central processing unit), whereas the decentralized THP (dTHP) considers that $\mathbf{C}$ is included at the receivers.

Our proposed (TH-SP) attempts to completely remove the MUI. We implement it by applying an LQ decomposition on the equivalent channel estimate $\bar{\mathbf{G}}^{\text{T}}$, i.e., $\bar{\mathbf{G}}^{\text{T}}=\bar{\mathbf{L}}\bar{\mathbf{Q}}$, where $\bar{\mathbf{L}} \in \mathbb{C}^{K \times K}$ and $\bar{\mathbf{Q}} \in \mathbb{C}^{K \times N}$. %\textcolor{red}{please check the matrix dimensions. In eq. (4) below, you say F=Q but in the para above F is NxN. So, that means Q should also be NxN. But then it does not match LQ decomposition.} 
Denote the $(n,k)$-th element of the matrix $\bar{\mathbf{L}}$ by $\hat{l}_{n,k}$. Then, the respective three THP filters are
\begin{align}
\mathbf{F}&=\bar{\mathbf{Q}}^{H},\\
\mathbf{C}&=\text{diag}\left(\bar{l}_{1,1},\bar{l}_{2,2},\cdots,\bar{l}_{N,N}\right),\\
\mathbf{B}^{\left(\text{c}\right)}&=\bar{\mathbf{L}}\mathbf{C}, \\\quad\mathbf{B}^{\left(\text{d}\right)}&=\mathbf{C}\bar{\mathbf{L}},
\end{align}
where $\mathbf{B}^{\left(\text{c}\right)}$ and $\mathbf{B}^{\left(\text{c}\right)}$ denote the feedback filters for the centralized and decentralized architectures, respectively. 

Denote the coefficients of the feedback filter by $b_{n,k}$ and the symbols after feedback processing by $\Breve{s}_k$. Then, the feedback filter subtracts the interference from previous symbols as %\textcolor{red}{you need to first define $\Breve{s}_i$ used on the RHS below}
\begin{equation}
    \Breve{s}_k=s_k-\sum_{i=1}^{k-1}b_{k,i}\Breve{s}_i.
\end{equation}
The feedback filter amplifies the power of the transmitted signal. Therefore, a modulo operation is introduced to reduce the power of the transmitted signal as
%The symbols are mapped inside the modulation's boundary \textcolor{red}{unclear what is boundary here} using a modulo operation as
\begin{equation}
    \mathcal{M}\left(\Breve{s}_k\right)=\Breve{s}_k-\left\lfloor\frac{\text{Re}\left(\Breve{s}_k\right)}{\lambda}+\frac{1}{2}\right\rfloor\lambda-j\left\lfloor\frac{\text{Im}\left(\Breve{s}_k\right)}{\lambda}+\frac{1}{2}\right\rfloor\lambda,
\end{equation}
where $\text{Re}(\cdot)$ ($\text{Im}(\cdot)$) is the real (imaginary) part of its complex argument and the parameter $\lambda$ depends on the modulation alphabet and the power allocation scheme. Some common values of $\lambda$ when employing symbols with unit variance are $\lambda=2\sqrt{2}$ and $\lambda=4\sqrt{10}/5$ for QPSK and 16-QAM, respectively. 

Unlike linear precoders \cite{wence,gbd,siprec,siprec2,wlbd,rmmseprec,cqabd,rsbd,bbprec,zcprec}, THP introduces power and modulo losses in the system. The former comes from the energy difference between the original constellation and the transmitted symbols after precoding. The latter is caused by the modulo operation. Both losses can be neglected for analysis purposes and for moderate and large modulation sizes \cite{Zu2014,rsthp}.% \textcolor{red}{need references for these claims}

The modulo operation is modeled as the addition of a perturbation vector $\mathbf{d}\in\mathbb{C}^{K\times 1}$ to the transmitted symbols $\mathbf{s}$. On the other hand, the feedback processing is implemented through the inversion of the matrix $\mathbf{B}$. Thus, the vector of symbols after feedback processing $\Breve{\mathbf{s}} \in \mathbb{C}^{K\times1}$ is %\textcolor{red}{vectors breve s and d below have not been defined before}
\begin{align}           
\label{eq:breves}
\Breve{\mathbf{s}}=&\mathbf{B}^{-1}\left(\mathbf{s}+\mathbf{d}\right)\nonumber\\
=&\mathbf{B}^{-1}\mathbf{v}.
\end{align}
Therefore, the receive signal vectors for the centralized and decentralized structures are, respectively, %\textcolor{red}{check the dimensions. are these kx1 like eq. (1)?}\textcolor{blue}{Yes, they are $k\times 1$ vectors.}
\begin{equation}
\label{eq:yc1}
    \mathbf{y}^{\left(\text{c}\right)}=\frac{1}{\beta^{\left(\text{c}\right)}}\left(\mathbf{G}^{\text{T}}\beta^{\left(\text{c}\right)}\mathbf{F}\mathbf{C}\Breve{\mathbf{s}}+\mathbf{n}\right),
\end{equation}
and
\begin{equation}
\label{eq:yd1}
    \mathbf{y}^{\left(\text{d}\right)}=\frac{1}{\beta^{\left(\text{d}\right)}}\mathbf{C}\left(\mathbf{G}^{\text{T}}\beta^{\left(\text{d}\right)}\mathbf{F}\Breve{\mathbf{s}}+\mathbf{n}\right), 
\end{equation}
%where $\mathbf{\mathbf{y}^{\left(\text{c}\right)}}$ and $\mathbf{\mathbf{y}^{\left(\text{d}\right)}}$ denote the received vector when employing the centralized and the decentralized structure, respectively. 
where the parameters $\beta^{\left(\text{c}\right)}$ ($\beta^{\left(\text{d}\right)}$) represent scaling factor of the centralized (decentralized) structure introduced to fulfill the transmit power constraint and defined as
\begin{align}
&\beta^{\left(\text{c}\right)}\approx\sqrt{\frac{P_t}{K}}, &\beta^{\left(\text{d}\right)}\approx\sqrt{\frac{P_t}{\sum\limits_{k=1}^{K}\left(1/\bar{l}_{k,k}^2\right)}}.
\end{align}

Using the channel relation $\mathbf{G}^{\text{T}}=\frac{1}{\tau}\left(\bar{\mathbf{G}}^{\text{T}}-\tilde{\mathbf{G}}^{\text{T}}\right)$, where $\tau=\sqrt{1+\sigma_e^2}$, and substituting \eqref{eq:breves} in \eqref{eq:yc1} and \eqref{eq:yd1} yield, respectively,
\begin{align}
    \mathbf{y}^{\left(\text{c}\right)}=&\frac{1}{\tau}\mathbf{v}-\frac{1}{\tau}\tilde{\mathbf{G}}^{\text{T}}\mathbf{F}\mathbf{C}\mathbf{B}^{\left(\text{c}\right)^{-1}}\mathbf{v}+\frac{1}{\beta^{\left(\text{c}\right)}}\mathbf{n},\\
    \mathbf{y}^{\left(\text{d}\right)}=&\frac{1}{\tau}\mathbf{v}-\frac{1}{\tau}\mathbf{C}\tilde{\mathbf{G}}^{\text{T}}\mathbf{F}\mathbf{B}^{\left(\text{d}\right)^{-1}}\mathbf{v}+\frac{1}{\beta^{\left(\text{d}\right)}}\mathbf{C}\mathbf{n},
\end{align}
It follows that the received signal at user $k$ is
\begin{align}
    y^{\left(\text{c}\right)}_k=&\frac{1}{\tau}v_k-\frac{1}{\tau}\tilde{\mathbf{g}}^{\text{T}}_k\sum\limits_{i=1}^{K}v_i\mathbf{p}_i^{\left(\text{c}\right)}+\frac{1}{\beta^{\left(\text{c}\right)}}n_k,\\
    y^{\left(\text{d}\right)}_i=&\frac{1}{\tau}v_k-\frac{c_{k,k}}{\tau}\tilde{\mathbf{g}}^{\text{T}}_k\sum\limits_{i=1}^{K}v_i\mathbf{p}_i^{\left(\text{d}\right)}+\frac{c_{k,k}}{\beta^{\left(\text{d}\right)}}n_k,
\end{align}
where, to simplify the notations, we have substituted the matrices $\mathbf{P}^{\left(\text{c}\right)}=\mathbf{F}\mathbf{C}\mathbf{B}^{\left(\text{c}\right)^{-1}}\in \mathbb{C}^{N\times K}$ and $\mathbf{P}^{\left(\text{d}\right)}=\mathbf{F}\mathbf{B}^{\left(\text{d}\right)^{-1}}\in \mathbb{C}^{N\times K}$, whose $i$-th columns are $\mathbf{p}^{\left(\text{c}\right)}_i$ and $\mathbf{p}^{\left(\text{d}\right)}_i$, %denote the $i$-th column of matrix $\mathbf{P}^{\left(\text{c}\right)}$ and $\mathbf{P}^{\left(\text{d}\right)}$, 
respectively.

\subsection{Cluster-based TH precoders}
Denote the $K$ clusters of usersby $\mathcal{P}_k$, $k=1,\cdots,K$. While the user $k$ is always included in $\mathcal{P}_k$, the user $i$, $i\neq k$ is included in $\mathcal{P}_k$ if at least $N_a$ antennas provide service to user $i$ and all other users in  $\mathcal{P}_k$. Then, define the user selection matrix $\mathbf{U}_k \in \mathbb{R}^{\lvert\mathcal{P}_k\rvert\times K}$, where $\lvert\mathcal{P}_k\rvert$ is the cardinality of the set $\mathcal{P}_k$ and the $j$-th row of $\mathbf{U}_k$ is $\mathbf{u}_{j,k}$. In particular, $\mathbf{u}_{1,k}$ contains zeros in all positions except in the $l$-th, where $l$ is the $j$-th lowest index in $\mathcal{P}_k$. Similarly, the second row $\mathbf{u}_{2,k}$ contains a one at the $j$-th position, where $j$ is the second lowest index in $\mathcal{P}_k$ and all other coefficients are equal to zero. The subsequent rows of $\mathbf{U}_k$ are defined similarly. 

The reduced channel matrix is $\bar{\mathbf{G}}^{\text{T}}_k=\mathbf{U}_k\bar{\mathbf{G}}^{\text{T}}\in\mathbb{C}^{\lvert\mathcal{P}_k\rvert\times N}$, which is used to compute the TH precoder with reduced dimensions (THP-RD). Applying an LQ decomposition over the reduced channel matrix, i.e. $\bar{\mathbf{G}}_k^{\text{T}}=\bar{\mathbf{L}}_k\bar{\mathbf{Q}}_k$, where $\bar{\mathbf{L}}_k\in \mathbb{C}^{\lvert\mathcal{P}_k\rvert \times \lvert\mathcal{P}_k\rvert}$ and $\bar{\mathbf{Q}}_k\in \mathbb{C}^{\lvert\mathcal{P}_k\rvert \times N}$, produces the three THP filters as %\textcolor{red}{again, check and specify the dimensions}
\begin{align}
\mathbf{F}_k&=\bar{\mathbf{Q}}_k^{H},\\
\mathbf{C}_k&=\text{diag}\left(\bar{l}_{1,1},\bar{l}_{2,2},\cdots,\bar{l}_{\lvert\mathcal{P}_k\rvert,\lvert\mathcal{P}_k\rvert}\right),\\
\mathbf{B}_k^{\left(\text{c}\right)}&=\bar{\mathbf{L}}_k\mathbf{C}_k, \\\mathbf{B}_k^{\left(\text{d}\right)}&=\mathbf{C}_k\bar{\mathbf{L}}_k.
\end{align}

The set $\mathcal{P}_k$ is associated to $\bar{\mathbf{G}}^{\text{T}}_k$ and to the decoding of the information of user $k$ but the channel matrix $\bar{\mathbf{G}}^{\text{T}}_k$ has reduced dimensions. %\textcolor{red}{how can a set has reduced dimensions? do you mean to say the set has fewer elements? if yes, then where do you define the set with more elements?}\textcolor{blue}{Thanks for pointing this out. The channel matrix has reduced dimensions and not the set. I have included this clarification to avoid any further confusion.} 
Therefore, we need an index mapping to find the correct precoder. Denote this index by $q$ such that $\mathbf{u}_{q,k}$ contains a one in its $k$-th entry. It follows that the $q$-th column should be employed in the precoders denoted by $\mathbf{P}^{\left(\text{cTHP-RD}\right)}= [\mathbf{p}_1^{\left(\text{c}\right)'}\ldots \mathbf{p}_k^{\left(\text{c}\right)'} \ldots \mathbf{p}_{K}^{\left(\text{c}\right)'} ]$ and $\mathbf{P}^{\left(\text{dTHP-RD}\right)}= [\mathbf{p}_1^{\left(\text{d}\right)'}\ldots \mathbf{p}_k^{\left(\text{d}\right)'} \ldots \mathbf{p}_{K}^{\left(\text{d}\right)'} ]$ for the cTHP and dTHP structures, respectively. Then, the $k$-th columns of the respective precoding matrices are 
\begin{align}
\mathbf{p}^{\left(c\right)'}_k=&\left[\mathbf{F}_k\mathbf{C}_k\mathbf{B}^{\left(c\right)^{-1}}_k\right]_q,\\
    \mathbf{p}^{\left(d\right)'}_k=&\left[\mathbf{F}_k\mathbf{B}^{\left(d\right)^{-1}}_k\right]_q.
\end{align}

\section{Sum-rate performance}
To evaluate the proposed nonlinear schemes,  we employ the ergodic sum-rate (ESR) defined as
\begin{equation}
    S_r=\mathbb{E}\left[\sum_{k=1}^{K}\bar{R}_k\right],
\end{equation}
where $\bar{R}_k=\mathbb{E}\left[R_k|\hat{\mathbf{G}}\right]$ is the average rate and $R_k$ is the instantaneous rate of the $k$-th user. The rate $\bar{R}_k$ averages out the effects of the imperfect CSIT because the instantaneous rates are not achievable. %The average rate is equal to $\bar{R}_k=\mathbb{E}\left[R_k|\hat{\mathbf{G}}\right]$, where $R_k$ stands for the instantaneous rate of the $k$-th user.
Considering Gaussian codebooks, the instantaneous rate is 
\begin{equation}
    R_k=\log_2\left(1+\gamma_k\right),\label{general instantaneous rate}
\end{equation}
where $\gamma_k$ is the signal-to-interference-plus-noise ratio (SINR) at user $k$. 

Denote the SINR for the centralized and decentralized structures by $\gamma_k^{\left(\text{c}\right)}$ and $\gamma_k^{\left(\text{d}\right)}$, respectively. Then, depending on the specific THP structure used, we employ $\gamma_k^{\left(\text{c}\right)}$ or $\gamma_k^{\left(\text{d}\right)}$ in \eqref{general instantaneous rate} to obtain the instantaneous rate. To compute the SINR, we obtain the mean powers of the received signal at user $k$ for centralized and decentralized structures as, respectively,
\begin{align}
    \mathbb{E}\left[\lvert y^{\left(\text{c}\right)}_k\rvert^2\right]=&\frac{1}{\tau^2}+\frac{1}{\tau^2}\sum\limits_{\substack{i=1\\i\neq k}}^K\lvert\tilde{\mathbf{g}}_k^{\text{T}}\mathbf{p}_i^{\left(\text{c}\right)}\rvert^2+\frac{1}{\beta^{\left(\text{c}\right)}}\sigma_n^2+\frac{1}{\tau^2}d_g^{\left(\text{c}\right)},
\end{align}
and
\begin{align}
    \mathbb{E}\left[\lvert y^{\left(\text{d}\right)}_k\rvert^2\right]=&\frac{1}{\tau^2}+\frac{c_{k,k}^2}{\tau^2}\sum\limits_{\substack{i=1\\i\neq k}}^K\lvert\tilde{\mathbf{g}}_k^{\text{T}}\mathbf{p}_i^{\left(\text{d}\right)}\rvert^2+\frac{c_{k,k}^2}{\beta^{\left(\text{d}\right)}}\sigma_n^2+\frac{1}{\tau^2}d_g^{\left(\text{d}\right)},
\end{align} 
where $d_g^{\left(\text{c}\right)}=\lvert\tilde{\mathbf{g}}_k\mathbf{p}^{\left(\text{c}\right)}_k\rvert^2-2\text{Re}{\left(\tilde{\mathbf{g}}_k\mathbf{p}^{\left(\text{c}\right)}_k\right)}$ %\textcolor{red}{is $\Re$ for real part? If yes, then the notation is different than used in eq. (9)} 
and $d_g^{\left(\text{d}\right)}=c_{k,k}^2\lvert\tilde{\mathbf{g}}_k\mathbf{p}^{\left(\text{d}\right)}_k\rvert^2-2\text{Re}{\left(c_{k,k}\tilde{\mathbf{g}}_k\mathbf{p}^{\left(\text{d}\right)}_k\right)}$. This yields
\begin{align}    
\gamma_k^{\left(\text{c}\right)}=&\frac{1}{d_g^{\left(\text{c}\right)}+\sum\limits_{\substack{i=1\\ i\neq k}}^K \left\lvert\tilde{\mathbf{g}}_k^{\textrm{T}}\mathbf{p}_i\right\rvert^2+\frac{\tau^2}{\beta^{\left(c\right)^2}}\sigma_n^2},
\end{align}
and
\begin{align}    
\gamma_k^{\left(\text{d}\right)}=&\frac{1}{d_g^{\left(\text{d}\right)}+c_{k,k}^2\sum\limits_{\substack{i=1\\i\neq k}}^K\lvert\tilde{\mathbf{g}}_k^{\text{T}}\mathbf{p}_i^{\left(\text{d}\right)}\rvert^2+\frac{c_{k,k}^2\tau^2}{\beta^{\left(\text{d}\right)}}\sigma_n^2}.\label{SINR general precoder}
\end{align}
%\textcolor{red}{the gamma used in eq. (25) is without any superscripts. You should write somewhere that the ESR and instantaneous rates are also with superscripts. Otherwise, there is no direct substitution of the eqns. above in (25)}

\section{Numerical Experiments}
We assess the performance of the proposed TH precoders via numerical experiments. Throughout the experiments, the large scale fading coefficients are set to
\begin{equation}
     \zeta_{k,n}=P_{k,n}\cdot 10^{\frac{\sigma^{\left(\textrm{s}\right)}z_{k,n}}{10}},
 \end{equation}
 where $P_{k,n}$ is the path loss and the scalar $10^{\frac{\sigma^{\left(\textrm{s}\right)}z_{k,n}}{10}}$ include the shadowing effect with standard deviation $\sigma^{\left(\textrm{s}\right)}=8$. The random variable $z_{k,n}$ follows Gaussian distribution with zero mean and unit variance. The path loss was calculated using a three-slope model as\par\noindent\small
 \begin{align}
     P_{k,n}=\begin{cases}
  -L-35\log_{10}\left(d_{k,n}\right), & \text{$d_{k,n}>d_1$} \\
  -L-15\log_{10}\left(d_1\right)-20\log_{10}\left(d_{k,n}\right), & \text{$d_0< d_{k,n}\leq d_1$}\\
    -L-15\log_{10}\left(d_1\right)-20\log_{10}\left(d_0\right), & \text{otherwise,}
\end{cases}
 \end{align}\normalsize
 where $d_{k,n}$ is the distance between the $n$-th AP and the $k$-th user, $d_1=50$ m, $d_0= 10$ m, and the attenuation $L$ is \par\noindent %\small
 \begin{align}
 L=&46.3+33.9\log_{10}\left(f\right)-13.82\log_{10}\left(h_{\textrm{AP}}\right)\nonumber\\
     &-\left(1.1\log_{10}\left(f\right)-0.7\right)h_u+\left(1.56\log_{10}\left(f\right)-0.8\right),
 \end{align}\normalsize
 where $h_{\textrm{AP}}=15$ m and $h_{u}=1.65$ m are the positions of the APs and UEs above the ground, respectively. We consider a frequency of $f= 1900$ MHz. The noise variance is 
  \begin{equation}
     \sigma_n^2=T_o k_B B N_f,
 \end{equation}
 where $T_o=290$ K is the noise temperature, $k_B=1.381\times 10^{-23}$ J/K is the Boltzmann constant, $B=50$ MHz is the bandwidth and $N_f=10$ dB is the noise figure. The signal-to-noise ratio (SNR) %\textcolor{red}{is this gamma? if yes, then you described gamma as SINR, not SNR in the previous section}\textcolor{blue}{No, gamma denotes the SINR and is different from the SNR introduced here.}
 is \par\noindent
 \begin{equation}
     \text{SNR}=\frac{P_{t}\textrm{Tr}\left(\mathbf{G}^{\text{T}}\mathbf{G}^{*}\right)}{N K \sigma_n^2},
 \end{equation}\normalsize
where $\textrm{Tr}(\cdot)$ is the trace of its matrix argument.

For all experiments, we have $128$ APs randomly distributed over a square with side equal to $20$ km. The APs serve a total of $24$ users, which are geographically distributed. We considered a total of 10,000 channel realizations to compute the ESR. Specifically, we employed $100$ channel estimates and, for each channel estimate, we considered $100$ error matrices. It follows that the average rate was computed with $100$ error matrices.  %\textcolor{red}{so which metric uses 10k realizations?}
  
\begin{figure}[t]
\begin{center}
\includegraphics[width=0.995\columnwidth]{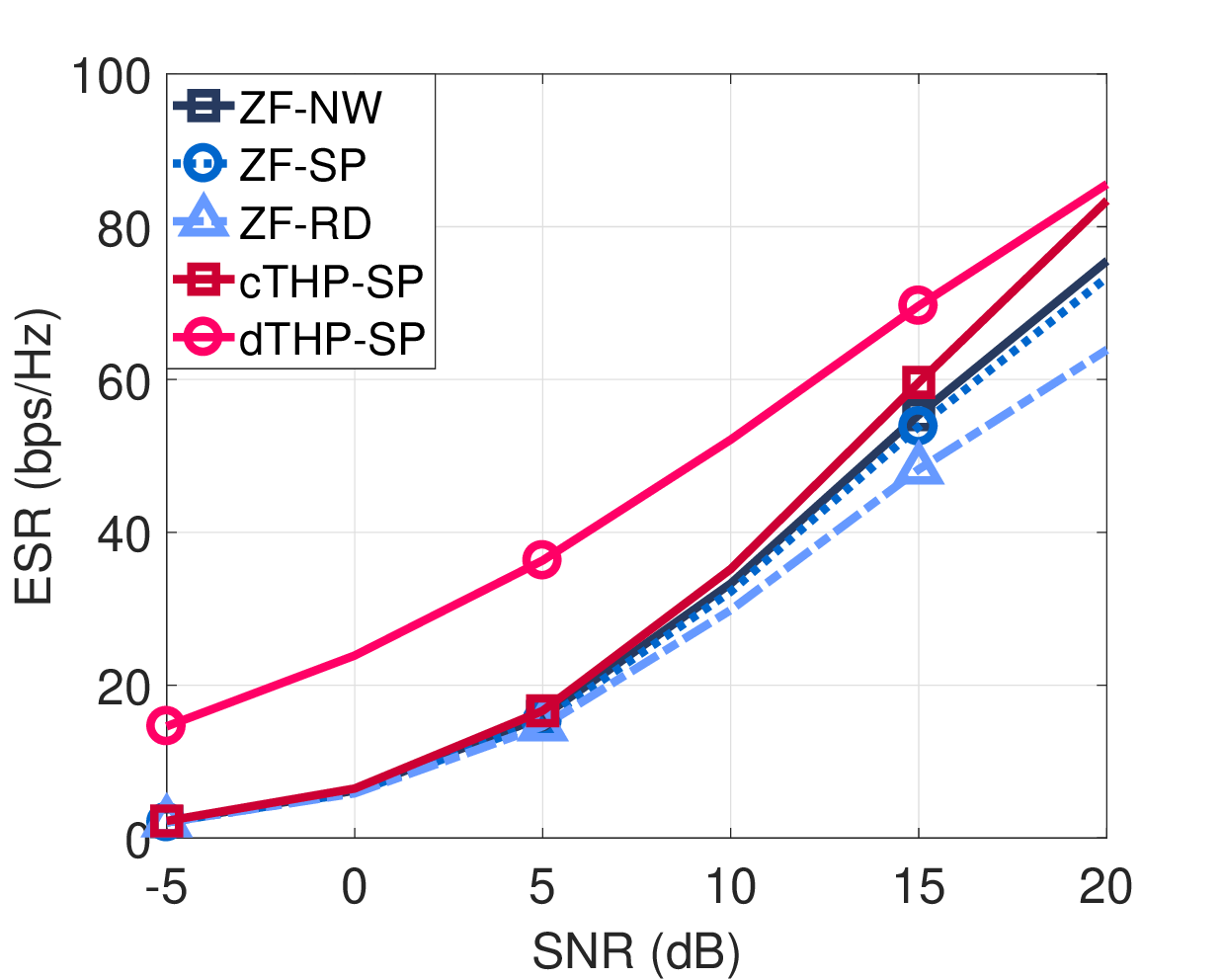}
\vspace{0.1em}
\caption{Sum-rate performance of various precoders versus SNR. Here, $N=128$, $K=24$, $|\mathcal{A}_k|=24$, $|\mathcal{P}_k|=10$, $\sigma_{e}^2=0.01$. }
\label{Fig1}
\end{center}
\end{figure}

We first compare the ESR of the proposed precoders with their linear counterparts. We consider that the error in the channel coefficient estimate has a variance of $0.01$. Fig.~\ref{Fig1} shows the sum-rate performance of the proposed nonlinear precoders against their conventional linear counterparts. The dTHP with sparse channel estimate or ``dTHP-SP'' performs the best, even better than the linear ZF precoder that employs all the APs (ZF-NW).%with complete \textcolor{red}{you mean `perfect'?} channel estimate (ZF-NW).
 
\begin{figure}[t]
\begin{center}
\includegraphics[width=0.995\columnwidth]{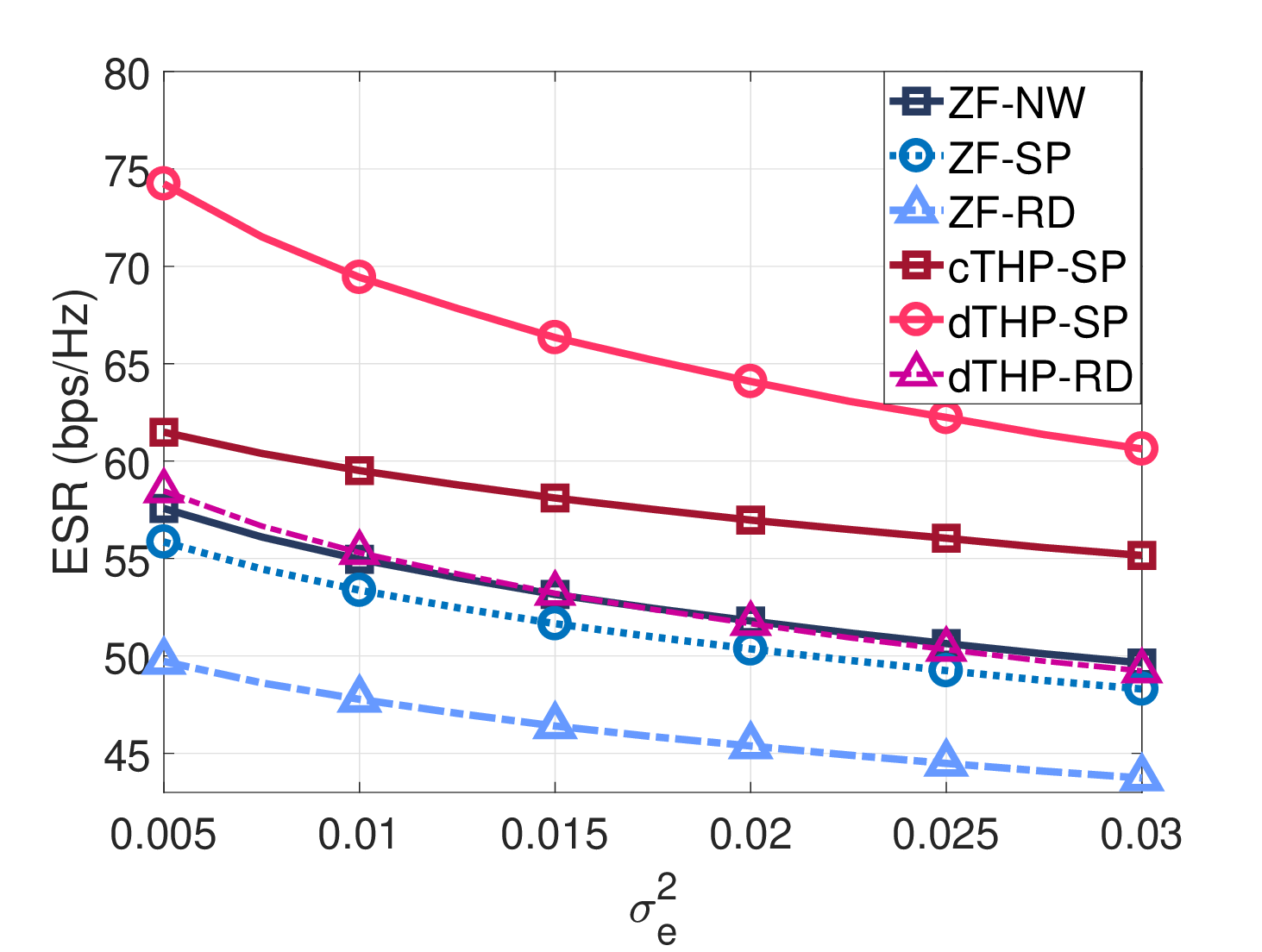}
\vspace{0.1em}
\caption{{Sum-rate performance of precoders versus CSIT quality. Here, $N=128$, $K=24$, $|\mathcal{A}_k|=24$, $|\mathcal{P}_k|=10$, $\textrm{SNR}=15 ~\textrm{dB}$.}}
\label{Fig3}
\end{center}
\end{figure}

In the second experiment, we assessed the sum-rate performance at SNR$=15$ dB with respect to CSIT quality (Fig.~\ref{Fig3}). The proposed dTHP with reduced dimensions or ``dTHP-RD'' outperforms the ZF-RD precoder. The RD precoding techniques have reduced computational complexity than the corresponding SP precoders. We observe that our proposed nonlinear cluster-based precoders generally yield better ESR than their linear counterparts.

\section{Summary}
We proposed clustered nonlinear precoders based on the noninear THP algorithm. Our proposed THP-SP reduces the signaling load and the THP-RD additionally lowers the computational complexity at the expense of performance. Note that the reduction of the computational complexity is critical for practical applications. Numerical experiments showed that the proposed cluster-based nonlinear precoders yield better performance and robustness against CSIT uncertainties than the conventional linear precoders.

\bibliographystyle{IEEEtran}
\bibliography{SubsetBib}

\begin{thebibliography}{10}
\providecommand{\url}[1]{#1}
\csname url@samestyle\endcsname
\providecommand{\newblock}{\relax}
\providecommand{\bibinfo}[2]{#2}
\providecommand{\BIBentrySTDinterwordspacing}{\spaceskip=0pt\relax}
\providecommand{\BIBentryALTinterwordstretchfactor}{4}
\providecommand{\BIBentryALTinterwordspacing}{\spaceskip=\fontdimen2\font plus
\BIBentryALTinterwordstretchfactor\fontdimen3\font minus \fontdimen4\font\relax}
\providecommand{\BIBforeignlanguage}[2]{{%
\expandafter\ifx\csname l@#1\endcsname\relax
\typeout{** WARNING: IEEEtran.bst: No hyphenation pattern has been}%
\typeout{** loaded for the language `#1'. Using the pattern for}%
\typeout{** the default language instead.}%
\else
\language=\csname l@#1\endcsname
\fi
#2}}
\providecommand{\BIBdecl}{\relax}
\BIBdecl

\bibitem{Tataria2021}
H.~Tataria, M.~Shafi, A.~F. Molisch, M.~Dohler, H.~Sjöland, and F.~Tufvesson, ``{6G} wireless systems: Vision, requirements, challenges, insights, and opportunities,'' \emph{Proceedings of the IEEE}, vol. 109, no.~7, pp. 1166--1199, 2021.

\bibitem{Giordani2020}
M.~Giordani, M.~Polese, M.~Mezzavilla, S.~Rangan, and M.~Zorzi, ``Toward {6G} networks: Use cases and technologies,'' \emph{IEEE Communications Magazine}, vol.~58, no.~3, pp. 55--61, 2020.

\bibitem{mmimo}
R.~C. de~Lamare, ``Massive {MIMO} systems: {S}ignal processing challenges and future trends,'' \emph{URSI Radio Science Bulletin}, vol. 2013, no. 347, pp. 8--20, 2013.

\bibitem{wence}
W.~Zhang, H.~Ren, C.~Pan, M.~Chen, R.~C. de~Lamare, B.~Du, and J.~Dai, ``Large-scale antenna systems with ul/dl hardware mismatch: Achievable rates analysis and calibration,'' \emph{IEEE Transactions on Communications}, vol.~63, no.~4, pp. 1216--1229, 2015.

\bibitem{Ammar2022}
H.~A. Ammar, R.~Adve, S.~Shahbazpanahi, G.~Boudreau, and K.~V. Srinivas, ``User-centric cell-free massive {MIMO} networks: {A} survey of opportunities, challenges and solutions,'' \emph{IEEE Communications Surveys \& Tutorials}, vol.~24, no.~1, pp. 611--652, 2022.

\bibitem{Elhoushy2021}
S.~Elhoushy, M.~Ibrahim, and W.~Hamouda, ``Cell-free massive {MIMO}: {A} survey,'' \emph{IEEE Communications Surveys \& Tutorials}, vol.~24, no.~1, pp. 492--523, 2022.

\bibitem{Attarifar2020}
M.~Attarifar, A.~Abbasfar, and A.~Lozano, ``Subset {MMSE} receivers for cell-free networks,'' \emph{IEEE Transactions on Wireless Communications}, vol.~19, no.~6, pp. 4183--4194, 2020.

\bibitem{Yang2018}
H.~Yang and T.~L. Marzetta, ``Energy efficiency of massive {MIMO}: {C}ell-free vs. cellular,'' in \emph{IEEE Vehicular Technology Conference - Spring}, 2018.

\bibitem{Elhoushy2021b}
S.~Elhoushy and W.~Hamouda, ``Towards high data rates in dynamic environments using hybrid cell-free massive {MIMO}/small-cell system,'' \emph{IEEE Wireless Communications Letters}, vol.~10, no.~2, pp. 201--205, 2021.

\bibitem{Ngo2018}
H.~Q. Ngo, L.-N. Tran, T.~Q. Duong, M.~Matthaiou, and E.~G. Larsson, ``On the total energy efficiency of cell-free massive {MIMO},'' \emph{IEEE Transactions on Green Communications and Networking}, vol.~2, no.~1, pp. 25--39, 2018.

\bibitem{Zhang2019}
J.~Zhang, S.~Chen, Y.~Lin, J.~Zheng, B.~Ai, and L.~Hanzo, ``Cell-free massive {MIMO}: A new next-generation paradigm,'' \emph{IEEE Access}, vol.~7, pp. 99\,878--99\,888, 2019.

\bibitem{Jin2021}
S.-N. Jin, D.-W. Yue, and H.~H. Nguyen, ``Spectral and energy efficiency in cell-free massive {MIMO} systems over correlated {R}ician fading,'' \emph{IEEE Systems Journal}, vol.~15, no.~2, pp. 2822--2833, 2021.

\bibitem{Nayebi2017}
E.~Nayebi, A.~Ashikhmin, T.~L. Marzetta, H.~Yang, and B.~D. Rao, ``Precoding and power optimization in cell-free massive {MIMO} systems,'' \emph{IEEE Transactions on Wireless Communications}, vol.~16, no.~7, pp. 4445--4459, 2017.

\bibitem{Bjoernson2020}
E.~Björnson and L.~Sanguinetti, ``Making cell-free massive {MIMO} competitive with {MMSE} processing and centralized implementation,'' \emph{IEEE Transactions on Wireless Communications}, vol.~19, no.~1, pp. 77--90, 2020.

\bibitem{Zu2014}
K.~Zu, R.~C. de~Lamare, and M.~Haardt, ``Multi-branch {T}omlinson-{H}arashima precoding design for {MU-MIMO} systems: {T}heory and algorithms,'' \emph{IEEE Transactions on Communications}, vol.~62, no.~3, pp. 939--951, 2014.

\bibitem{rmbthp}
L.~Zhang, Y.~Cai, R.~C. de~Lamare, and M.~Zhao, ``Robust multibranch {T}omlinson–{H}arashima precoding design in amplify-and-forward {MIMO} relay systems,'' \emph{IEEE Transactions on Communications}, vol.~62, no.~10, pp. 3476--3490, 2014.

\bibitem{rsthp}
A.~R. Flores, R.~C. De~Lamare, and B.~Clerckx, ``Tomlinson-{H}arashima precoded rate-splitting with stream combiners for {MU-MIMO} systems,'' \emph{IEEE Transactions on Communications}, vol.~69, no.~6, pp. 3833--3845, 2021.

\bibitem{Joham2005}
M.~Joham, W.~Utschick, and J.~Nossek, ``Linear transmit processing in {MIMO} communications systems,'' \emph{IEEE Transactions on Signal Processing}, vol.~53, no.~8, pp. 2700--2712, 2005.

\bibitem{lrcc}
H.~Ruan and R.~C. de~Lamare, ``Distributed robust beamforming based on low-rank and cross-correlation techniques: Design and analysis,'' \emph{IEEE Transactions on Signal Processing}, vol.~67, no.~24, pp. 6411--6423, 2019.

\bibitem{Palhares2020}
V.~M. Palhares, R.~C. de~Lamare, A.~R. Flores, and L.~T. Landau, ``Iterative {AP} selection, {MMSE} precoding and power allocation in cell-free massive {MIMO} systems,'' \emph{IET Communications}, vol.~14, no.~22, pp. 3996--4006, 2020.

\bibitem{rsrbd}
A.~R. Flores, R.~C. de~Lamare, and B.~Clerckx, ``Linear precoding and stream combining for rate splitting in multiuser {MIMO} systems,'' \emph{IEEE Communications Letters}, vol.~24, no.~4, pp. 890--894, 2020.

\bibitem{cesg}
S.~Mashdour, R.~C. de~Lamare, and J.~P. S.~H. Lima, ``Enhanced subset greedy multiuser scheduling in clustered cell-free massive mimo systems,'' \emph{IEEE Communications Letters}, vol.~27, no.~2, pp. 610--614, 2023.

\bibitem{cfrs}
A.~R. Flores, R.~C. de~Lamare, and K.~V. Mishra, ``Clustered cell-free multi-user multiple-antenna systems with rate-splitting: Precoder design and power allocation,'' \emph{IEEE Transactions on Communications}, vol.~71, no.~10, pp. 5920--5934, 2023.

\bibitem{Ngo2017}
H.~Q. Ngo, A.~Ashikhmin, H.~Yang, E.~G. Larsson, and T.~L. Marzetta, ``Cell-free massive {MIMO} versus small cells,'' \emph{IEEE Transactions on Wireless Communications}, vol.~16, no.~3, pp. 1834--1850, 2017.

\bibitem{Nguyen2017}
L.~D. Nguyen, T.~Q. Duong, H.~Q. Ngo, and K.~Tourki, ``Energy efficiency in cell-free massive {MIMO} with zero-forcing precoding design,'' \emph{IEEE Communications Letters}, vol.~21, no.~8, pp. 1871--1874, 2017.

\bibitem{Buzzi2020}
S.~Buzzi, C.~D’Andrea, A.~Zappone, and C.~D’Elia, ``User-centric {5G} cellular networks: Resource allocation and comparison with the cell-free massive {MIMO} approach,'' \emph{IEEE Transactions on Wireless Communications}, vol.~19, no.~2, pp. 1250--1264, 2020.

\bibitem{Bjoernson2020a}
E.~Björnson and L.~Sanguinetti, ``Scalable cell-free massive {MIMO} systems,'' \emph{IEEE Transactions on Communications}, vol.~68, no.~7, pp. 4247--4261, 2020.

\bibitem{Lozano2021}
M.~M. Mojahedian and A.~Lozano, ``Subset regularized zero-forcing precoders for cell-free {C-RAN}s,'' in \emph{European Signal Processing Conference}, 2021, pp. 915--919.

\bibitem{albreem2021overview}
M.~A. Albreem, A.~H. Al~Habbash, A.~M. Abu-Hudrouss, and S.~S. Ikki, ``Overview of precoding techniques for massive {MIMO},'' \emph{IEEE Access}, vol.~9, pp. 60\,764--60\,801, 2021.

\bibitem{Fischer2002}
R.~Fischer, C.~Windpassinger, A.~Lampe, and J.~Huber, ``Tomlinson-{H}arashima precoding in space-time transmission for low-rate backward channel,'' in \emph{International Zurich Seminar on Broadband Communications Access - Transmission - Networking}, 2002, pp. 7--7.

\bibitem{spa}
R.~C. De~Lamare and R.~Sampaio-Neto, ``Minimum mean-squared error iterative successive parallel arbitrated decision feedback detectors for ds-cdma systems,'' \emph{IEEE Transactions on Communications}, vol.~56, no.~5, pp. 778--789, 2008.

\bibitem{Vu2007}
M.~Vu and A.~Paulraj, ``{MIMO} wireless linear precoding,'' \emph{IEEE Signal Processing Magazine}, vol.~24, no.~5, pp. 86--105, 2007.

\bibitem{chen2019low}
Y.~Chen, ``Low complexity precoding schemes for massive {MIMO} systems,'' Ph.D. dissertation, Newcastle University, 2019.

\bibitem{gbd}
K.~Zu, R.~C. de~Lamare, and M.~Haardt, ``Generalized design of low-complexity block diagonalization type precoding algorithms for multiuser mimo systems,'' \emph{IEEE Transactions on Communications}, vol.~61, no.~10, pp. 4232--4242, 2013.

\bibitem{siprec}
Y.~Cai, R.~C.~d. Lamare, and R.~Fa, ``Switched interleaving techniques with limited feedback for interference mitigation in ds-cdma systems,'' \emph{IEEE Transactions on Communications}, vol.~59, no.~7, pp. 1946--1956, 2011.

\bibitem{siprec2}
Y.~Cai, R.~C. de~Lamare, and D.~Le~Ruyet, ``Transmit processing techniques based on switched interleaving and limited feedback for interference mitigation in multiantenna mc-cdma systems,'' \emph{IEEE Transactions on Vehicular Technology}, vol.~60, no.~4, pp. 1559--1570, 2011.

\bibitem{wlbd}
W.~Zhang, R.~C. de~Lamare, C.~Pan, M.~Chen, J.~Dai, B.~Wu, and X.~Bao, ``Widely linear precoding for large-scale mimo with iqi: Algorithms and performance analysis,'' \emph{IEEE Transactions on Wireless Communications}, vol.~16, no.~5, pp. 3298--3312, 2017.

\bibitem{rmmseprec}
Y.~Cai, R.~C. de~Lamare, L.-L. Yang, and M.~Zhao, ``Robust mmse precoding based on switched relaying and side information for multiuser mimo relay systems,'' \emph{IEEE Transactions on Vehicular Technology}, vol.~64, no.~12, pp. 5677--5687, 2015.

\bibitem{cqabd}
S.~F.~B. Pinto and R.~C. de~Lamare, ``Block diagonalization precoding and power allocation for multiple-antenna systems with coarsely quantized signals,'' \emph{IEEE Transactions on Communications}, vol.~69, no.~10, pp. 6793--6807, 2021.

\bibitem{rsbd}
A.~R. Flores, R.~C. de~Lamare, and B.~Clerckx, ``Linear precoding and stream combining for rate splitting in multiuser mimo systems,'' \emph{IEEE Communications Letters}, vol.~24, no.~4, pp. 890--894, 2020.

\bibitem{bbprec}
L.~T.~N. Landau and R.~C. de~Lamare, ``Branch-and-bound precoding for multiuser mimo systems with 1-bit quantization,'' \emph{IEEE Wireless Communications Letters}, vol.~6, no.~6, pp. 770--773, 2017.

\bibitem{zcprec}
D.~M.~V. Melo, L.~T.~N. Landau, R.~C. de~Lamare, P.~F. Neuhaus, and G.~P. Fettweis, ``Zero-crossing precoding techniques for channels with 1-bit temporal oversampling adcs,'' \emph{IEEE Transactions on Wireless Communications}, vol.~22, no.~8, pp. 5321--5336, 2023.

\end{thebibliography}

\end{document}